\documentclass[12pt]{iopart}
\usepackage{graphics}  
\begin{document}
%
\title[A ${\cal PT}$ symmetric Natanzon-class potential]{An 
exactly solvable ${\cal PT}$ symmetric potential 
from the Natanzon class}

\author{G. L\'evai\dag\footnote[3]{E-mail: levai@atomki.hu},  
        A. Sinha\ddag\  and P. Roy$\|$ }

\address{\dag\ Institute of Nuclear Research of the Hungarian
         Academy of Sciences, \\
         PO Box 51, H--4001 Debrecen, Hungary}
\address{\ddag\ Department of Applied Mathematics, 
         Calcutta University, 92 APC Road, Kolkata 700009, India}
\address{$\|$\ Physics and Applied Mathematics Unit, Indian 
         Statistical Institute, Kolkata 700035, India}

\begin{abstract}
The ${\cal PT}$ symmetric version of the generalised Ginocchio
potential, a member of the general exactly solvable Natanzon potential 
class is analysed and its properties are compared with those of 
${\cal PT}$ symmetric potentials from the more restricted
shape-invariant class. It is found that the ${\cal PT}$ symmetric 
generalised Ginocchio potential has a number of properties in common 
with the latter potentials: it can be generated by an imaginary
coordinate shift $x\rightarrow x+{\rm i}\varepsilon$; its states are 
characterised by the quasi-parity quantum number; the spontaneous
breakdown of ${\cal PT}$ symmetry occurs at the same time for all the
energy levels; and it has two supersymmetric partners which cease to be 
${\cal PT}$ symmetric when the ${\cal PT}$ symmetry of the original 
potential is spontaneously broken. 
\end{abstract}


\maketitle

\section{Introduction}

Exactly solvable examples played an important role in the understanding of 
${\cal PT}$ symmetric quantum mechanics and its unusual features. The 
first ${\cal PT}$ symmetric potentials, i.e. potentials which are invariant 
under the {\it simultaneous} action of the ${\cal P}$ space- and ${\cal T}$ 
time reflection operations were found numerically \cite{bb98}. The most 
surprising result was that these one-dimensional complex potentials 
possessed {\it real} energy eigenvalues, which however, turned pairwise 
into complex conjugated pairs as some potential parameter was tuned. This 
mechanism was interpreted as the spontaneous breakdown of ${\cal PT}$ 
symmetry, since the potential remained ${\cal PT}$ invariant throughout, 
but the eigenfunctions associated with the discrete spectrum lost this 
property as the spectrum turned into complex gradually with the tuning 
of the potential parameter. Further examples have been found in numerical 
\cite{num}, semiclassical \cite{semi} and perturbative 
\cite{pert} studies, but a number of quasi-exactly solvable (QES) 
\cite{qespt} and exactly solvable 
\cite{mzho,glmz00,expt} ${\cal PT}$ symmetric 
potentials have also been identified. These latter potentials were 
analogues of Hermitian exactly solvable potentials with the property that 
their real and imaginary components were even and odd functions of the 
coordinate $x$, respectively. 

More recently ${\cal PT}$ symmetric problems have been analysed in 
terms of pseudo-Hermiticity \cite{mo02}, and their unusual features have 
been interpreted in terms of this more general context. A Hamiltonian is 
said to be $\eta$-pseudo-Hermitian if there exists a linear, invertible, 
Hermitian operator for which $H^{\dag}=\eta H\eta^{-1}$ holds. It has 
been shown that for systems with Hamiltonians of the type 
$H=p^2+V(x)$ ${\cal PT}$ symmetry is equivalent with 
${\cal P}$-pseudo-Hermiticity. With other choices of $\eta$ further 
complexified Hamiltonians can be generated, which might also have 
real energy eigenvalues but do not fulfill ${\cal PT}$ symmetry 
\cite{bq00,ahmed01}. Furthermore, with $\eta=1$ $\eta$-pseudo-Hermiticity 
reduces to conventional Hermiticity. 

Here we restrict our analysis to ${\cal PT}$ symmetric problems, and 
in particular, to exactly solvable ones. A number of peculiar features of 
${\cal PT}$ symmetric potentials became apparent only during the analysis 
of these potentials. 

\begin{itemize}
\item
It has been observed that several exactly solvable ${\cal PT}$ symmetric 
potentials possess {\it two} sets of normalisable solutions 
\cite{mzho,bq00,jpa01} in the sense that there can be two normalisable 
states with the same principal quantum number $n$. The second set of 
solutions can appear in several ways. For potentials which are singular 
at the origin the problem can be redefined on various trajectories of the 
complex $x$ plane such that the integration path avoids the origin and the 
solutions remain asymptotically normalisable. (This latter feature is 
similar to some numerically solvable ${\cal PT}$ symmetric problems 
\cite{bb98}.) A special case of this scenario is obtained when an 
{\it imaginary} coordinate shift $x\rightarrow x+{\rm i}\varepsilon$ is 
employed \cite{glmz00}. The advantage of this scenario is that the 
discussion of the ${\cal PT}$ symmetric potential remains rather similar 
to that of its Hermitian counterpart, and by suitable and straightforward 
modification of the formalism it can also be interpreted as a 
conventional complex potential defined on the $x$ axis. This imaginary 
coordinate shift can be employed \cite{glmz00} to all the 
shape-invariant \cite{si} potentials (e.g. the ${\cal PT}$ symmetric 
harmonic oscillator \cite{mzho} or the generalised P\"oschl--Teller 
potential \cite{jpa02a}) with the exception of the 
${\cal PT}$ symmetric Morse \cite{mzmorse} and Coulomb \cite{mzgl00} 
potentials. The cancellation of the singularity then regularizes the 
solution which would be irregular at the origin in the Hermitian setting. 
A different mechanism appears for potentials which are not singular in 
their Hermitian version, such as the ${\cal PT}$ symmetric Scarf II 
potential, which is defined on the full $x$ axis. In this case the 
second set of normalisable solutions originates from states which have 
complex eigen-energy in the Hermitian case, but which turn into 
normalisable states with real energy when  the potential is forced to 
become ${\cal PT}$ symmetric and the ${\cal PT}$ symmetry is not broken 
spontaneously \cite{jpa01,ahptbr,jpa02a,pla02}. The two set of solutions 
are distinguished by the quasi-parity quantum number \cite{quasip}.

\item
In the process of generating the spontaneous breakdown of ${\cal PT}$ 
symmetry by tuning the potential parameters it was found that the 
pairwise merging of the energy eigevalues and their re-emergence as 
complex conjugated pairs occurs at the {\it same} value of the potential 
parameter \cite{mpla01a,ahptbr}. In other words, the spontaneous 
breakdown of ${\cal PT}$ 
symmetry is realised suddenly in the case of shape-invariant potentials, 
as opposed to a gradual process observed in the case of numerical 
examples \cite{bb98,mpla01b}.

\item 
For certain exactly solvable (and shape-invariant) examples, such as 
the ${\cal PT}$ symmetric harmonic oscillator \cite{ptho} and the 
Scarf II potential \cite{jpa02b} it was found that there are {\it two} 
`fermionic' SUSY partners of the original 
`bosonic' potential, and they are distinguished by the quasi-parity 
quantum number carried by the `bosonic' bound states. 
(In the latter case this has also been found in other realisations of 
SUSYQM \cite{bmq02}.) This doubling of the partner potentials is an 
obvious consequence of the fact that there are {\it two} nodeless 
normalisable solutions corresponding to the `ground state' in the 
two segments of the spectrum with quasi-parity $q=+1$ and $-1$. 
Furthermore, it was also established that the 
`fermionic' partner potentials are ${\cal PT}$ symmetric themselves 
too, in case the `bosonic' potential has unbroken ${\cal PT}$ 
symmetry, while they cease to be ${\cal PT}$ symmetric if the 
${\cal PT}$ symmetry of the `bosonic' potential is spontaneously
broken. 

\end{itemize}

The peculiar features mentioned above have been observed until now 
for the ${\cal PT}$ symmetric version of shape-invariant potentials, 
while potentials beyond this class often behaved in a different way. 
This naturally raises the question whether these features also 
characterise non-shape-invariant, but exactly solvable examples. 
Natural candidates for this analysis are Natanzon-class potentials 
\cite{natanzon} which have the property that their bound-state 
solutions are written in terms of a {\it single} hypergeometric or 
confluent hypergeometric function. This potential class depends on six 
parameters, but it is prohibitively complicated in its general form, 
so its subclasses with two to four parameters have been analysed in 
detail until now \cite{gin84,gin85,bs90,imppots,dkv,rrzl01}. 

Perhaps the most well-known member of the Natanzon-class is the Ginocchio 
potential, which has a one-dimesional version defined on the $x$ axis 
\cite{gin84} and a radial one, which has an $r^{-2}$-like singularity at 
the origin \cite{gin85}. An important feature of this potential is that 
for a special choice of a potential parameter it reduces to a 
shape-invariant potential, namely to the P\"oschl--Teller hole (in one 
dimension) and to the generalised P\"oschl--Teller potential (in the 
radial case). The one-dimensional version of the Ginocchio potential  has 
been analysed in an algebraic framework, and an su(1,1) algebra has been 
associated with it \cite{ail85}, the discrete non-unitary irreducible 
representations of which correspond to resonances in the transmission 
coefficients. (These states have been identified as quasi-bound states 
in an independent study \cite{ahgin}.) Furthermore, it was also shown 
that this algebra reduces in 
two different shape-invariant limits to an su(1,1) potential algebra and 
to an su(2) spectrum generating algebra \cite{gos96}. The 
phase-equivalent supersymmetric partners of the generalised Ginocchio 
potential have also been derived in a completely analytic form 
\cite{jpa97}. 

Although the generalised Ginocchio potential is an 
`implicit' potential, i.e. the $z(r)$ function which is used to 
transform the Schr\"odinger equation into the differential equation 
of the hypergeometric function $F(a,b;c;z)$ 
is known only in an implicit form 
as $r(z)$, nevertheless, similarly to a number of other `implicit' 
potentials \cite{imppots}, this does not restrict the applicability of 
the formulae, because $V(r)$ and the wavefunctions can be determined to 
any desired accuracy, and all the calculations involving these quantities 
(matrix elements, etc.) can be evaluated analytically \cite{jpa97}. 
Furthermore, we shall see that the imaginary coordinate shift which is 
essential to impose ${\cal PT}$ symmetry on the generalised Ginocchio 
potential can also be implemented without complications. 

We note that a Natanzon-class potential, the so-called DKV potential 
has already been analysed in the ${\cal PT}$ symmetric setting by the  
point canonical transformation of a shape-invariant potential \cite{zlrr01}, 
but it was found that it has to be defined on a curved integration path. 
Nevertheless, it also showed similarities with ${\cal PT}$ symmetric 
shape-invariant potentials, as its spectrum was also richer that that 
of its Hermitian counterpart. 

In section \ref{ggp} we present the Hermitian version of the generalised 
Ginocchio potential for reference, and in section \ref{ptggp} we 
construct its ${\cal PT}$ symmetric version. Section \ref{sspt} deals 
with the supersymmetric aspects of this potential, while in section 
\ref{conc} a summary of the results is presented.

\section{The generalised Ginocchio potential}
\label{ggp}

The first version of the Ginocchio potential was introduced 
as a one-dimensional quantum mechanical problem which is 
symmetric with respect to the $x\rightarrow -x$ transformation 
\cite{gin84}. Later it was generalised to a radial problem \cite{gin85}, 
which also contains an $r^{-2}$-like singular term at the 
origin. (This latter version of the potential also allows a 
particular functional form of an effective mass, but it can be 
reduced to a constant value by setting one of the parameters ($a$) 
to zero.) 
Following the notation of Ref. \cite{jpa97} 
we define the generalised Ginocchio potential as
\begin{eqnarray}
V(r)&=&-\frac{\gamma^4(s(s+1) + 1-\gamma^2)}{\gamma^2 + \sinh^2 u} 
+\gamma^4\lambda(\lambda-1)\frac{\coth^2 u}{\gamma^2 + \sinh^2 u}
\nonumber\\
&&-\frac{3 \gamma^4(\gamma^2-1)(3\gamma^2-1)}{4(\gamma^2+\sinh^2 u)^2}
+\frac{5\gamma^6 (\gamma^2-1)^2}{4(\gamma^2+\sinh^2 u)^3} \ ,
\label{ginpot}
\end{eqnarray}
where we changed the notation of Ref. \cite{gin85} to make it more 
suitable for our purposes. This form can be obtained from the 
original formulae by setting $a=0$, $\alpha_l=\lambda-{1\over
2}$, $\nu_l=s$, $\beta_{nl}=\mu$, $\lambda=\gamma$ and $y=\sinh
u(\gamma^2 +\sinh^2 u)^{-{1\over 2}}$. 

The (generalised) Ginocchio potential is an example for 
`implicit' potentials, because it is expressed in terms of 
a function $u(r)$ which is known only in the implicit $r(u)$ 
form: 
\begin{eqnarray}
r={1\over \gamma^2}&&\Big[\tanh^{-1}\Big(
   (\gamma^2 +\sinh^2 u)^{-{1\over 2}} \sinh u \Big) \Big. 
\nonumber\\
&&\Big.
+(\gamma^2-1)^{1\over 2}\tan^{-1}\Big((\gamma^2-1)^{1\over 2}
  (\gamma^2 +\sinh^2 u)^{-{1\over 2}} \sinh u \Big)\Big] \ .
\label{ginru}
\end{eqnarray}
$r$ can take values from the positive half axis, which is 
mapped by the monotonously increasing implicit $u(r)$ function 
onto itself. 
This function is, actually, the solution of an ordinary 
first-order differential equation 
\begin{equation}
{{\rm d}u\over {\rm d}r}={\gamma^2 \cosh u \over 
(\gamma^2 + \sinh^2 u)^{1\over 2}}
\label{dudr}
\end{equation}
defining a variable transformation connecting the Schr\"odinger
equation with the differential equation of the Jacobi
(and Gegenbauer) polynomials \cite{as70}. It can be seen from
Eqs. (\ref{ginru}) and (\ref{dudr}) that $u(r)$ behaves approximately as
$\gamma r$ near the origin, and as $\gamma^2 r$ for large values
of $r$. In the $\gamma\rightarrow 1$ limit $u$ becomes identical
with $r$, and (\ref{ginpot}) reduces to the generalised 
P\"oschl--Teller potential.  

Bound states are located at 
\begin{equation}
E_n=-\gamma^4\mu^2_n \ ,
\label{ginen}
\end{equation}
where $n$ varies from 0 to $n_{\rm max}$ defined below and
\begin{equation}
\fl
\mu_n={1\over \gamma^2}\Big[ -\Big( 2n+\lambda+\frac{1}{2}\Big)
+\Big[ \Big( 2n+\lambda+\frac{1}{2}\Big)^2(1-\gamma^2)
+\gamma^2\Big( s+\frac{1}{2}\Big)^2 \Big]^{1\over 2}\Big] \ .
\label{ginmu}
\end{equation}
All the terms in (\ref{ginpot}) are finite at the origin, with the
exception of the last one, which shows $r^{-2}$-like singularity
there, and can be considered either as an approximation of the 
centrifugal term with $l=\lambda-1$
($\lambda$ integer), or as a part of a singular potential 
with arbitrary $l\neq\lambda-1$. Setting $\lambda=1$ or 0 we get the
`simple' Ginocchio potential \cite{gin84} defined on the line. 

The bound-state wavefunctions are expressed in terms of Jacobi
polynomials 
\begin{equation}
\fl
\psi_n(r)={\cal N}_n (\gamma^2+ \sinh^2 u)^{1\over 4}
(\sinh u)^{\lambda}(\cosh u)^{-\mu_n-\lambda-\frac{1}{2}} 
P^{(\mu_n , \lambda-{1\over 2})}_n(2\tanh^2 u -1)
\label{ginbound}
\end{equation}
which reduce to Gegenbauer polynomials \cite{as70} for $\lambda=1$. 
The normalisation is given by 
\begin{equation}
{\cal N}_n = \Big[ {2\gamma^2 n!\ \Gamma(\mu_n+\lambda+n+
\frac{1}{2})\mu_n(\mu_n + \lambda + 2n + \frac{1}{2}) \over
\Gamma(\mu_n + n + 1) \Gamma(\lambda + n + \frac{1}{2}) (\mu_n 
\gamma^2 + \lambda + 2n + \frac{1}{2}) }\Big]^{\frac{1}{2}}.
\label{ginnorm}
\end{equation}
Considering that the $r\rightarrow\infty$ asymptotical limit 
corresponds to $u\rightarrow\infty$ (see Eq. (\ref{ginru})), the
wavefunctions become zero asymptotically if $\mu_n>0$ holds.
Applying this condition to Eq. (\ref{ginmu}) we find that the number of
bound states is set by $n_{\rm max}<{1\over 2}(s-\lambda)$.

\section{${\cal PT}$ symmetrisation of the generalised Ginocchio 
potential}
\label{ptggp}

\begin{figure}[h]
\begin{center}
\resizebox*{16cm}{!}{\includegraphics*[0cm,9cm][22cm,18cm]{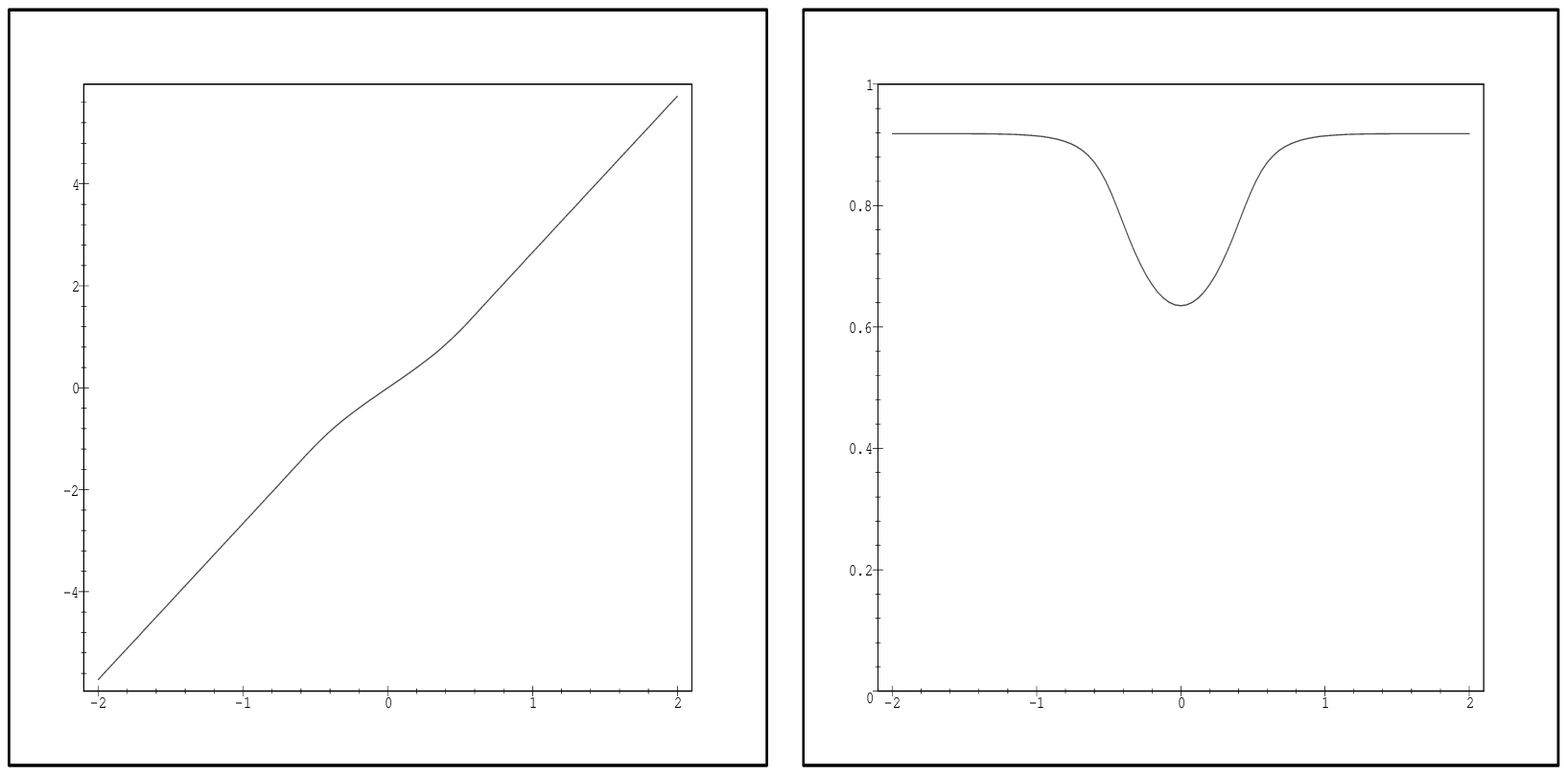}}
\vspace{.5cm}
\caption{The real (left panel) and imaginary (right panel) component of 
the $u(x)$ function for $\gamma=1.75$ and $\varepsilon=0.3$. Note the 
different vertical scales. 
}
\label{uxie}
\end{center}
\end{figure}

The first step in the ${\cal PT}$ symmetrisation of the generalised 
Ginocchio potential is performing the imaginary coordinate shift 
which allows its extension to the full $x$ axis by cancelling the 
singularity at the origin. This imaginary coordinate shift is a 
constant of integration from (\ref{dudr}), and it modifies (\ref{ginru}) 
such that $r\rightarrow x+{\rm i\varepsilon}$. Here we also switched to 
$x$ instead of $r$ to indicate that the original radial potential is 
extended also to the negative $x$ axis, following the standard 
treatment of ${\cal PT}$ symmetric potentials. 
Similarly to the Hermitian case, the variable transformation is 
determined by an implicit formula, 
\begin{eqnarray}
x+{\rm i\varepsilon}={1\over \gamma^2}&&\Big[\tanh^{-1}\Big(
   (\gamma^2 +\sinh^2 u)^{-{1\over 2}} \sinh u \Big) \Big. 
\nonumber\\
&&\Big.
+(\gamma^2-1)^{1\over 2}\tan^{-1}\Big((\gamma^2-1)^{1\over 2}
  (\gamma^2 +\sinh^2 u)^{-{1\over 2}} \sinh u \Big)\Big] 
\label{ginrupt}
\end{eqnarray}
however, now $u$ takes on complex values. 
Figure \ref{uxie} shows the $u(x)$ function for a particular value of 
$\gamma$ and $\varepsilon$. This function varies smoothly and it is odd 
under the ${\cal PT}$ transformation: ${\cal PT}u(x)=-u(x)$, i.e. 
its real and imaginary components are odd and even functions of $x$, 
respectively. Asymptotically the relation 
$u(x)\rightarrow_{x\rightarrow\pm\infty} \gamma^2 (x+{\rm i}\varepsilon)$
holds, while near $x=0$ there is a `kink' in both the real and the 
imaginary component of $u(x)$. 

The ${\cal PT}$ transform of $\sinh u(x+{\rm i}\varepsilon)$ is 
$-\sinh u(x+{\rm i}\varepsilon)$ (which can be seen analytically too by 
series expansion), 
and this also determines the ${\cal PT}$ transform of the potential 
(\ref{ginpot}). In particular, we may note that it contains 
$\sinh u$ everywhere as $\sinh^2 u$ (including also the term with 
$\coth^2$), so (\ref{ginpot}) is ${\cal PT}$ symmetric if all the 
coupling coefficients are {\it real}. This restricts $\gamma^2$, 
$s(s+1)$ and $\lambda(\lambda-1)$ to real values. The latter two 
requirements allow the following values of $s$ and $\lambda$:
\begin{equation}
\fl
s=\cases{ {\rm real} & when $s(s+1)\ge -\frac{1}{4}$ \cr 
          -\frac{1}{2}+{\rm i}\sigma & when $s(s+1)\le -\frac{1}{4}$ }
\hskip .5cm
\lambda=\cases{ {\rm real} & when $\lambda(\lambda-1)\ge -\frac{1}{4}$ \cr 
        \frac{1}{2}+{\rm i}l & when $\lambda(\lambda-1)\le -\frac{1}{4}$ }
\label{slambda}
\end{equation}
We are going to discuss these possibilities later. 

Let us now 
analyse the two independent solutions of the generalised 
Ginocchio potential, written in the form hypergeometric 
functions. 
From among the possible linear combinations we chose the 
ones which have different behaviour at the origin in the 
$\varepsilon\rightarrow 0$ limit:
\begin{equation}
\fl
\psi_1(x)\sim (\gamma^2+\sinh^2 u)^{1/4}(\cosh u)^{a-b}
(\sinh u)^{a+b-c+\frac{1}{2}} F(a,a-c+1;a+b-c+1;-\sinh^2 u)\ ,
\label{hipg1}
\end{equation}
\begin{equation}
\fl
\psi_2(x)\sim (\gamma^2+\sinh^2 u)^{1/4}(\cosh u)^{a-b}
(\sinh u)^{c-a-b+\frac{1}{2}} F(1-b,c-b;c-a-b+1;-\sinh^2 u)\ ,
\label{hipg2}
\end{equation}
where $a$, $b$ and $c$ have to satisfy the following relations: 
\begin{equation}
\fl
c=1\pm \mu\hskip .7cm 
a+b-c=\pm(\lambda-\frac{1}{2})\hskip .7cm
a-b=\pm\left[(s+\frac{1}{2})^2-(\gamma^2-1)\mu^2\right]^{1/2}
\equiv\pm \omega\ .
\label{abc}
\end{equation}
With the conditions (\ref{abc}) the two independent solutions can be 
written as 
\begin{eqnarray}
\fl
\psi_1(x)\sim && (\gamma^2+\sinh^2 u)^{1/4}(\cosh u)^{\pm\omega}
(\sinh u)^{\lambda} 
\nonumber\\
\fl
&& \times F(\frac{1}{2}(\mu+\lambda+\frac{1}{2}\pm\omega), 
\frac{1}{2}(-\mu+\lambda+\frac{1}{2}\pm\omega);
\lambda+\frac{1}{2};-\sinh^2 u)\ ,
\label{hipg1x}
\end{eqnarray}
\begin{eqnarray}
\fl
\psi_2(x)\sim && (\gamma^2+\sinh^2 u)^{1/4}(\cosh u)^{\pm\omega}
(\sinh u)^{1-\lambda} 
\nonumber\\
\fl
&& \times F(\frac{1}{2}(-\mu-\lambda+\frac{3}{2}\pm\omega), 
\frac{1}{2}(\mu-\lambda+\frac{3}{2}\pm\omega);
\frac{3}{2}-\lambda;-\sinh^2 u)\ . 
\label{hipg2x}
\end{eqnarray}
Note that the same two functions are obtained irrespective of the 
signs chosen in the first two equations in (\ref{abc}), while the 
sign of $\omega$ remains to be determined from the 
normalisability conditions of the wavefunctions. 

Up to this point the functions (\ref{hipg1x}) and 
(\ref{hipg2x}) supply the general solutions for the energy eigenvalue
$E=-\gamma^4\mu^2$. In order to obtain solutions belonging to discrete 
energy eigenvalues one has to set one of the first two arguments of the 
hypergeometric functions to the non-positive integer value $-n$,
reducing them  to Jacobi polynomials \cite{as70}. We find that in 
contrast with the Hermitian case, normalisable solutions can be 
obtained in {\it two} different ways, corresponding to the condition 
\begin{equation}
2n+1+\mu_{nq}+q(\lambda-\frac{1}{2})-\omega=0\ ,
\label{polcond}
\end{equation}
where $q=1$ and $q=-1$ holds for (\ref{hipg1x}) and (\ref{hipg2x}), 
respectively. In this case the two solutions can be written in 
a compact form as 
\begin{eqnarray}
\psi_{n q}(x)\sim && (\gamma^2+\sinh^2 u)^{1/4}
(\cosh u)^{-2n-1-\mu_{n q}-q(\lambda-\frac{1}{2})}
(\sinh u)^{\frac{1}{2}+q(\lambda-\frac{1}{2})}
\nonumber\\
&&\times 
P_n^{(q(\lambda-\frac{1}{2}), 
      -2n-1-\mu_{n q}-q(\lambda-\frac{1}{2}))}(\cosh(2u))\ .
\label{psinq}
\end{eqnarray}
Here $n$ is the principal quantum number labelling the bound states
and $q$ is the quasi-parity $q=\pm 1$ \cite{quasip}. This quantum number
characterises the solutions of ${\cal PT}$ symmetric potentials, but the
potential itself does not depend on it. Its name originates from the
analysis of the ${\cal PT}$ symmetric version of the one-dimensional 
harmonic oscillator: in the Hermitian limit of this potential (i.e. 
for $\varepsilon\rightarrow 0$ it essentially reduces to the parity 
quantum number. 

We see that the two solutions are distinguished by the $q=\pm 1$ 
quasi-parity quantum number, similarly to potentials belonging to the 
shape-invariant class. Actually, the corresponding solutions 
of the ${\cal PT}$ symmetric generalised P\"oschl--Teller potential 
\cite{jpa02a} can be obtained by setting $\gamma=1$. 
It is also obvious that normalisability requires 
${\rm Re}(\mu_{n q})>0$. 

In order to obtain explicit expression for $\mu_{nq}$ one has to 
combine (\ref{polcond}) with (\ref{abc}) and to solve a quadratic 
algebraic equation for $\mu=\mu_{nq}$: 
\begin{equation}
\fl
\mu_{nq}=\frac{1}{\gamma^2}\left[-\left(2n+1+q(\lambda-\frac{1}{2})\right)
+\left[\gamma^2(s+\frac{1}{2})^2
+(1-\gamma^2)\left( 2n+1+q(\lambda-\frac{1}{2})\right)^2\right]^{1/2}
\right]
\label{munq}
\end{equation}
This expression recovers the corresponding formula for the Hermitian 
generalised Ginocchio potential for $q=1$. 

Similarly to the case of the Hermitian version of the generalised 
Ginocchio potential the energy eigenvalues are written as 
$E_{n q} =-\gamma^4\mu^2_{n q}$, 
and they are independent of $\varepsilon$. Actually, we find that for the 
$q=1$ choice the expressions for the Hermitian problem are recovered 
formally. However, the forthcoming analysis will show that despite 
the similar form, some quantities can be chosen complex for the 
${\cal PT}$ symmetric case. Before going on we may note that $\mu_{nq}$,
and consequently $E_{n q}$ depends on the $2n+1+q(\lambda-\frac{1}{2})$ 
combination, and this leads to a degeneracy between levels with $q=1$
and $q=-1$ whenever $\lambda$ is a real half-integer number.
Furthermore, for $\lambda=\frac{1}{2}$ states with opposite quasi-parity
and with the same $n$ become degenerate: in fact, this is the point
where the spontaneous breakdown of ${\cal PT}$ symmetry sets in if the 
$\lambda$ parameter is continued to complex values allowed by 
(\ref{slambda}).

Let us now analyse the conditions for having real and complex 
energy eigenvalues $E_{n q}$, which corresponds to inspecting the nature 
of $\mu_{n q}$ (\ref{munq}) in terms of the allowed values of $s$ and 
$\lambda$ displayed in (\ref{slambda}). These also have to be combined 
with the condition ${\rm Re}(\mu_{n q}) >0$ which guarantees 
normalisability of the solutions (\ref{psinq}). The key element of the 
analysis is the term containing the square root in (\ref{munq}), so 
it is useful to inspect separately the cases when it is real, imaginary 
or complex, which corresponds to $A\ge 0$, $A<0$ and complex $A$, where 
\begin{equation}
A\equiv \gamma^2(s+\frac{1}{2})^2
+(1-\gamma^2)\left( 2n+1+q(\lambda-\frac{1}{2})\right)^2\ .
\label{ginsq}
\end{equation}
We restrict our analysis to $\gamma^2>1$: the alternative 
choice, $\gamma^2<1$ would change the nature of the $r(u)$ function 
in (\ref{ginru}). We can note that 
$\lambda$ occurs everywhere in the combination $q(\lambda-\frac{1}{2})$, 
so when $\lambda$ is real, we can assume that $\lambda\ge \frac{1}{2}$, 
because the $\lambda\le \frac{1}{2}$ cases can be obtained simply by 
switching $q=+1$ to $q=-1$. Also, when $\lambda=\frac{1}{2}+{\rm i}l$, 
it is enough to assume $l>0$ for the same reason. 

\begin{itemize}

\item
$A\ge 0$. This can happen only if $\lambda$ is real, while from $A\ge 0$ 
in (\ref{ginsq}) and $\gamma^2>1$ it follows that $s$ also has 
to be real. Inspecting the allowed values of $n$ for various parameter 
domains we find the following. Normalizable states can be obtained 
for $\mu_{n q}$ in (\ref{munq}) when 
\begin{equation}
\fl
-\frac{1}{2}\left(1+q(\lambda-\frac{1}{2})+\left(
\frac{\gamma^2}{\gamma^2-1}\right)^{1/2}\vert s+\frac{1}{2}\vert\right)
\le n \le 
-\frac{1}{2}\left(1+q(\lambda-\frac{1}{2})-\vert s+\frac{1}{2}\vert\right)
\label{posap}
\end{equation}
holds. 
If the upper boundary of this domain is negative, then there are no 
normalizable solutions. This depends on the relative magnitude of 
$s$ and $\lambda$. 

\item
$A< 0$. Here again $\lambda$ has to be real, while $s$ can take both 
real and complex values allowed in (\ref{slambda}). The 
${\rm Re}(\mu_{n q})>0$ condition now reduces to 
$2n+1+q(\lambda-\frac{1}{2})<0$, which has to be combined with 
$A< 0$. The resulting condition is then 
\begin{equation}
n \le -\frac{1}{2}\left(1+q(\lambda-\frac{1}{2})+\left(
\frac{\gamma^2}{\gamma^2-1}\right)^{1/2}\vert s+\frac{1}{2}\vert\right)
\label{negasr}
\end{equation}
for real values of $s$ and 
\begin{equation}
n \le -\frac{1}{2}\left(1+q(\lambda-\frac{1}{2})\right)
\label{negasi}
\end{equation}
for $s=-\frac{1}{2}+{\rm i}\sigma$. Note that these conditions can be 
met only for $q=-1$ if $\lambda$ is large enough (and positive, as we 
assumed before). 

\item
$A$ is complex. For this $\lambda=\frac{1}{2}+{\rm i}l$ is required, 
while $s$ can be both real and $s=-\frac{1}{2}+{\rm i}\sigma$. 
This situation corresponds to the spontaneous breakdown of 
${\cal PT}$ symmetry, and the energy eigenvalues appear in 
complex conjugated pairs due to $(\mu_{n q})^*=\mu_{n -q}$, which 
leads to $(E_{n q})^*=E_{n -q}$. At the same time the 
${\rm Re}(\mu_{n q})>0$ condition turns out to be the same for 
$q=+1$ and $-1$, in accordance with the expectation that the number 
of normalisable states has to be the same for both quasi-parities. 
The detailed analysis is more complicated for complex values of 
$A$ (and $\lambda$) than for real $A$, so we can resort only to 
numerical calculations in this respect. The outcome depends on 
the relative magnitude of $\vert s+\frac{1}{2}\vert$ 
(which is $\vert\sigma\vert$ for complex values of $s$) and $l$.

\end{itemize}

We can now address the question whether with the tuning of the potential 
parameters the spontaneous breakdown of 
${\cal PT}$ symmetry  (i.e. the appearance of complex conjugate 
pairs of eigenvalues) happens at the same time for {\it all} the bound
states as in the case of shape-invariant potentials \cite{mpla01a}, or 
gradually, as for some non-shape-invariant potentials, such as the 
${\cal PT}$ symmetric square well \cite{mpla01b}. From the analysis 
above we find that when this mechanism
is realised via setting $\lambda$ to the complex value 
$\lambda=\frac{1}{2}+{\rm i}l$, then all the energy eigenvalues 
turn to complex at the same time. The spectrum can also be changed to 
complex by tuning $s$ from real to imaginary values and keeping
$\lambda$ real. Since $s$ is contained in the formulae in the
combination $s(s+1)$ or $(s+\frac{1}{2})^2=s(s+1)+\frac{1}{4}$ which 
is always real, this possibility is more limited. In this case again 
all the energy eigenvalues turn into complex at the same time, but 
the character of the potential also changes, as its leading term 
in (\ref{ginpot}) changes sign. In the Hermitian setting this would
correspond to replacing the potential well with a barrier, which
obviously changes the nature of the problem. Although in the ${\cal PT}$
symmetric version of (\ref{ginpot}) the situation is less transparent,
the complexification of the spectrum via tuning $s$ to complex values
and keeping $\lambda$ real is clearly different from the situation 
when $s$ is kept real and the spontaneous breakdown of ${\cal PT}$
symmetry is induced by tuning $\lambda$ to complex values. 

\begin{figure}[h]
\begin{center}
\resizebox*{16cm}{!}{\includegraphics*[0cm,9cm][22cm,18cm]{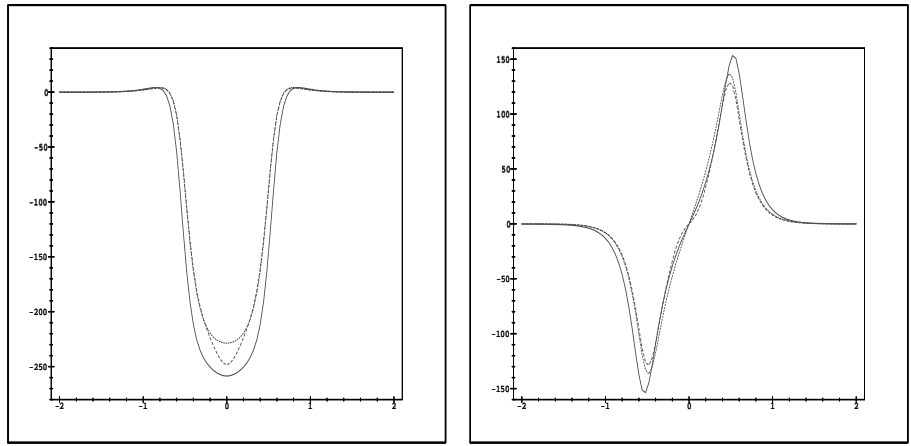}}
\vspace{.5cm}
\caption{The real (left panel) and imaginary (right panel) component of 
the potential (\ref{ginpot}) for $\varepsilon=0.3$, $\gamma=1.75$, 
$s=8.1$ and $\lambda=1.25$ (solid line) and its supersymmetric partners 
$V^{(+1)}_+(x)$ (dashed line) and $V^{(-1)}_+(x)$ (dotted line) in 
(\ref{ginpotpart}). Normalisable states of (\ref{ginpot}) are found at 
$E_{0\ +1}=-171.313$, $E_{1\ +1}=-106.160$, $E_{2\ +1}=-46.679$, 
$E_{3\ +1}=-5.666$; 
$E_{0\ -1}=-218.913$,$E_{1\ -1}=-154.978$, $E_{2\ -1}=-90.379$, 
$E_{3\ -1}=-33.993$ and $E_{4\ -1}=-1.061$. The spectrum of 
$V^{(q)}_+(x)$ is the same, with the exception of the $E_{0\ q}$ level, 
which is missing from its spectrum. 
}
\label{examp1}
\end{center}
\end{figure}

\begin{figure}[h]
\begin{center}
\resizebox*{16cm}{!}{\includegraphics*[0cm,9cm][22cm,18cm]{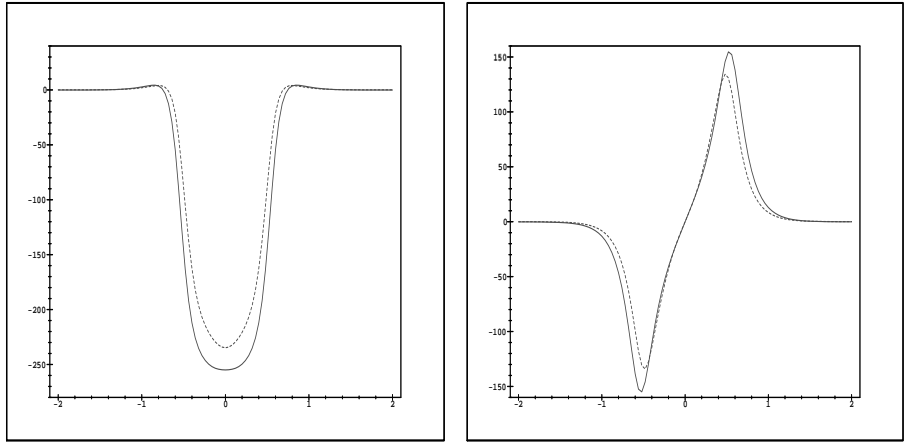}}
\vspace{.5cm}
\caption{The same as Figure \ref{examp1} with 
$\lambda=0.5$. Normalisable states are found at 
$E_{0\ +1}=E_{0\ -1}=-195.477$, $E_{1\ +1}=E_{1\ -1}=-130.419$, 
$E_{2\ +1}=E_{2\ -1}=-67.675$ and $E_{3\ +1}=E_{3\ -1}=-17.640$.  
The supersymmetric partners $V^{(+1)}_+(x)$ and $V^{(-1)}_+(x)$ coincide
in this case. 
}
\label{examp2}
\end{center}
\end{figure}

In Figures \ref{examp1}, \ref{examp2} and \ref{examp3} the real and 
imaginary components of (\ref{ginpot}) are plotted for fixed values 
of $\varepsilon$, $\gamma$ and $s$ and for various values of $\lambda$ 
corresponding to unbroken and spontaneously broken ${\cal PT}$ symmetry. 
The position of the energy eigenvalues are also indicated. 

A similar analysis can be performed for $\gamma^2<1$ too. In this case
the $s=-\frac{1}{2}+{\rm i}\sigma$ choice plays a more important role,
but otherwise the results are qualitatively the same. Note that in this
case $r(u)$ in (\ref{ginru}) changes, and this also modifies the nature
of the potential. 

Before closing this section we mention briefly some aspects of the
one-dimesional version of the Ginocchio potential \cite{gin84} which is
obtained from (\ref{ginpot}) by the $\lambda=0$ or 1 substitution. 
This limit is analogous to the one-dimensional version of the harmonic 
oscillator, which is also obtained from the radial harmonic oscillator
after cancelling the singular centrifugal term, allowing the 
extension of the potential to the full $x$ axis. Another similarity
between the two systems is that the two choices of $\lambda$ correspond
to the even and odd solutions, and this can clearly be seen from the
structure of the bound-state solutions (\ref{ginbound}), in which the
Jacobi polynomial reduces to an even and odd Gegenbauer polynomial for 
$\lambda=0$ and 1, respectively \cite{as70}. Losing the $\lambda$
parameter means that in the case of the ${\cal PT}$ symmetric 
one-dimensional Ginocchio potential the spontaneous breakdown of 
${\cal PT}$ symmetry cannot be implemented as in the general case. It is
also interesting to note that the quasi-parity quantum number occurs
only in the combination $q(\lambda-\frac{1}{2})$, which means that
the $q=+1$, $\lambda=0$ combination is equivalent with $q=-1$, 
$\lambda=1$, and $q=+1$, $\lambda=1$ is equivalent with $q=-1$, 
$\lambda=0$, and this reflects the relation of the quasi-parity quantum
number with ordinary parity, similarly to the case of the one-dimensional 
harmonic oscillator \cite{quasip}. It is also worthwhile to note that
the Hermitian version of the one-dimensional Ginocchio potential
possesses a number of complex-energy solutions (resonances) 
\cite{gin84}, and since the energy eigenvalues are not sensitive to the 
$\varepsilon$ parameter appearing in the complex coordinate shift in 
(\ref{ginrupt}), these remain unchanged after the ${\cal PT}$ 
symmetrisation of the potential. However, these are unbound solutions, 
so their character is different from that of the (normalisable) 
complex-energy solutions which appear when the ${\cal PT}$ symmetry is 
spontaneously broken.

\section{Supersymmetric aspects of the ${\cal PT}$ symmetric 
generalised Ginocchio potential}
\label{sspt}

According to Ref. \cite{jpa02b} the supersymmetric partner of a 
${\cal PT}$ symmetric potential depends on the quasi-parity $q$, and 
thus corresponds to {\it two distinct} potentials. For this, the partner
potentials have to be constructed by using a $q$-dependent factorization
energy such that \cite{jpa02b}
\begin{equation}
V^{(q)}_{\pm}(x)=U^{(q)}_{\pm}(x)+\epsilon^{(q)}\equiv 
[W^{(q)}(x)]^2 \pm\frac{{\rm d}W^{(q)}}{{\rm d}x}+\epsilon^{(q)}\ ,
\label{vpm}
\end{equation}
where $\epsilon^{(q)}=E^{(q)}_{0,-}$ is the ground-state energy of 
the `bosonic' potential , which, due to this construction is 
independent from $q$, i.e. $V^{(q)}_-(x)=V(x)$. The superpotential 
$W^{(q)}(x)$ is expressed in terms of the ground-state $(n=0)$ 
wavefunction of $V(x)$ (\ref{ginpot}) 
\begin{eqnarray}
\fl
W^{(q)}(x)&=&-\frac{\rm d}{{\rm d}x}\ln 
\psi_{0 q}(x)
\nonumber\\
\fl
&=&\frac{\gamma^2(\gamma^2-1)\sinh u}{2(\gamma^2+\sinh^2 u)^{3/2}}
+\frac{\gamma^2\mu_{0 q}\sinh u}{(\gamma^2+\sinh^2 u)^{1/2}}
-\frac{\gamma^2(\frac{1}{2}+q(\lambda-\frac{1}{2}))}{
(\gamma^2+\sinh^2 u)^{1/2}\sinh u}\ ,
\label{wgin}
\end{eqnarray}
and it clearly depends on $q$ explicitly (in third term) and implicitly  
via $\mu_{0 q}$ (in the second term). The 
`fermionic' partner potentials of the generalised Ginocchio potential 
contain the same terms as (\ref{ginpot}), but the coupling coefficients 
are different, and pick up $q$-dependence, as expected \cite{jpa02b}: 
\begin{equation}
\fl
V^{(q)}_+(x)=
-\frac{A_q}{\gamma^2 + \sinh^2 u} 
+B_q\frac{\coth^2 u}{\gamma^2 + \sinh^2 u}
+\frac{C_q}{(\gamma^2+\sinh^2 u)^2}
-\frac{7\gamma^6 (\gamma^2-1)^2}{4(\gamma^2+\sinh^2 u)^3} \ ,
\label{ginpotpart}
\end{equation}
\begin{eqnarray}
A_q&=&\gamma^4[s(s+1) + \gamma^2-2 -2\gamma^2\mu_{0 q}-q(2\lambda-1)]\ ,
\\
B_q&=&\gamma^4[\lambda(\lambda-1)+1+q(2\lambda-1)]\ ,
\\
C_q&=&\gamma^4(\gamma^2-1)\left(\frac{11\gamma^2-9}{4}-2\gamma^2\mu_{0 q}
-q(2\lambda-1) \right)
\end{eqnarray}

These partner potentials are ${\cal PT}$ symmetric if $\mu_{0 q}$ (and 
thus $\lambda$ too) are real, which, under rather general conditions, 
coincides with the requirement of the (unbroken) ${\cal PT}$ symmetry of 
(\ref{ginpot}) itself. This was the case for some ${\cal PT}$ 
symmetric shape-invariant potentials too \cite{jpa02b,ptho}. When 
$\lambda=\frac{1}{2}+{\rm i}l$, which happens when the ${\cal PT}$ 
symmetry of (\ref{ginpot}) is spontaneously broken, (\ref{ginpotpart})
ceases to be ${\cal PT}$ symmetric, which is again a result similar to
those obtained for shape-invariant potentials \cite{jpa02b,ptho}. 

\begin{figure}[h]
\begin{center}
\resizebox*{16cm}{!}{\includegraphics*[0cm,9cm][22cm,18cm]{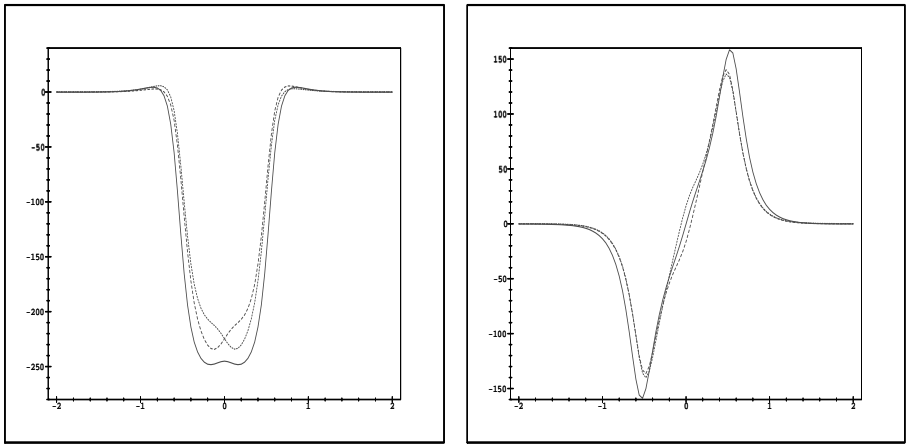}}
\vspace{.5cm}
\caption{The same as Figure \ref{examp1} with $\lambda=0.5+1.25{\rm i}$ 
corresponding to spontaneously broken ${\cal PT}$ symmetry. Normalisable 
states are found at 
$E_{0\ +1}=(E_{0\ -1})^*=-196.494+{\rm i}\ 40.038$, 
$E_{1\ +1}=(E_{1\ -1})^*=-130.023+{\rm i}\ 41.130$, 
$E_{2\ +1}=(E_{2\ -1})^*=-65.367+{\rm i}\ 37.105$ and 
$E_{3\ +1}=(E_{3\ -1})^*=-11.833+{\rm i}\ 25.161$.    
The supersymmetric partners $V^{(+1)}_+(x)$ and $V^{(-1)}_+(x)$ 
cease to be ${\cal PT}$ symmetric in this case. 
}
\label{examp3}
\end{center}
\end{figure}

Figures \ref{examp1}, \ref{examp2} and \ref{examp3} display also the 
`fermionic' partners of the respective `bosonic' potentials. 
Due to the SUSYQM construction the energy eigenvalues of the 
`fermionic' partners are the same with the exception that the levels 
with 
$n=0$ and $q=\pm 1$ are missing from the spectrum of $V^{(\pm 1)}_+(x)$.
The example in figure \ref{examp3} corresponds to the spontaneous
breakdown of the ${\cal PT}$ symmetry of the `bosonic' potential, and 
thus the ${\cal PT}$ symmetry of the `fermionic' potentials is 
manifestly broken. This is indicated by the fact that the real and 
imaginary component of the potential ceases to have definite parity 
under space reflexion. However, the two `fermionic' potentials are 
the ${\cal PT}$ transforms of each other: 
$V^{(+1)}_+(x)=[V^{(-1)}_+(-x)]^*$. Note that for 
$\lambda=\frac{1}{2}$, i.e. for the point of the spontaneous breakdown 
of ${\cal PT}$  symmetry the two `fermionic' partners with $q=+1$ 
and $q=-1$ coincide.

\section{Summary and conlcusions}
\label{conc}

We analysed a Natanzon-class potential, the generalised Ginocchio 
potential in a ${\cal PT}$ symmetric setting in order to explore 
similarities and differences with the more restricted shape-invariant 
potential class. This work was inspired by the fact that up to now 
the exactly solvable ${\cal PT}$ symmetric potentials were almost 
exclusively members of the shape-invariant class, and they showed 
marked differences compared to examples outside this class, e.g. 
those which have been solved numerically or had quasi-exactly solvable 
character. Our analysis showed that the ${\cal PT}$ symmetric generalised 
Ginocchio potential shares all the specific properties of shape-invariant 
potentials. In particular, its states can also be characterised by the 
quasi-parity quantum number, and the spontaneous breakdown of its 
${\cal PT}$ symmetry takes place suddenly, i.e. by tuning a potential 
parameter ($\lambda$) all its real energy eigenvalues turn into complex 
conjugate pairs at the same value of this parameter. 
These results seem to originate from the `robust' structure of the 
normalisable solutions of Natanzon-class potentials, which allows the 
implementation of ${\cal PT}$ symmetry to these technically non-trivial 
problems. 
Another similarity with the shape-invariant potentials is that the 
${\cal PT}$ symmetric generalised Ginocchio potential
has two `fermionic' supersymmetric partners (generated by 
eliminating the lowest state of the original `bosonic' potential 
with quasi-parity $q=+1$ and $-1$), and the partner potentials also 
possess ${\cal PT}$ symmetry as long as the ${\cal PT}$ symmetry of the 
`bosonic' potential is unbroken, but they cease to be ${\cal PT}$ 
symmetric when the ${\cal PT}$ symmetry of the `bosonic' potential is 
spontaneously broken. This seems to indicate that the properties thought 
to be specific to ${\cal PT}$ symmetric shape-invariant potentials might 
be valid to the much larger Natanzon potential class too, and perhaps 
also beyond that. Further studies should be made to check the validity 
of this conjecture.

\ack
This work was supported by the OTKA grant No. T031945 (Hungary)
and by the MTA--INSA (Hungarian--Indian) cooperation. 
A.S. acknowledges financial assistance from CSIR, INDIA.

\newpage

\end{document}